\documentclass[aps,pra,amsfonts,amssymb,amsmath,showpacs,
floatfix,nofootinbib,groupedaddress,superscriptaddress,citesort,twocolumn]{revtex4}

\usepackage{amssymb,amsmath,amsthm}

\theoremstyle{definition}

\theoremstyle{remark}

\usepackage{subfigure}
\usepackage{mathrsfs}
\usepackage{longtable}
\usepackage{amsfonts}
\usepackage{amstext}
\usepackage[usenames]{xcolor}
\usepackage[dvips]{graphicx}
\def\qed{\leavevmode\unskip\penalty9999 \hbox{}\nobreak\hfill
     \quad\hbox{\leavevmode  \hbox to.77778em{%
              \hfil\vrule   \vbox to.675em%
               {\hrule width.6em\vfil\hrule}\vrule\hfil}}
     \par\vskip3pt}

\begin{document}
\title{Characterizing  the  superposition of  arbitrary random quantum states and a known quantum state}

\author{Bo Li}
\email{libobeijing2008@163.com.}
\affiliation{School of Computer and Computing Science, Hangzhou City University, Hangzhou 310015, China}
\author{Xiao-Bin Liang}
\email{liangxiaobin2004@126.com.}
\affiliation{School of Mathematics and Computer science, Shangrao Normal University, Shangrao 334001, China}
\author{Shao-Ming Fei}
\email{feishm@cnu.edu.cn}
\affiliation{School of Mathematical Sciences, Capital Normal University, Beijing 100048, China}

\begin{abstract}
The superposition of states is one of the most fundamental issues in the quantum world. Generally there do not exist physical operations to superpose two unknown random states with nonzero probability. We investigate the superposition problem of unknown qubit states with respect to a known qubit state. It is shown that under trace-nonincreasing completely positive operations the superposable state sets are located in some circles on the Bloch sphere. Meanwhile, we prove that the quantum states in a circle on the Bloch sphere are superposable with respect to a known state. Finally, for the high-dimensional case, we illustrate that any superposition transformation protocols will violate the no-cloning principle for almost all the states. Our results also promote the understanding and applications of the superposition principle in view of quantum no-go theorems.
\end{abstract}

\pacs{03.67.-a, 03.65.Ud, 03.65.Yz}
\maketitle

\section{Introduction.}
The quantum state superposition is considered to be the essence of numerous nonclassical phenomena in quantum theory, such as quantum interference \cite{1,2}, quantum coherence, entanglement and other correlations \cite{3,4,5,6,Halder,Aubrun,MLHu}, with applications in quantum information processing like teleportation \cite{c}. It also gives rise to the great power of quantum computation in such as the Shor's decomposition algorithm \cite{7} or the Boson sampling \cite{8,H,Y}.
However, there is no quantum superposition type \emph{quantum adder}. Although the global phase of a quantum state is physically trivial, the different relative phases of the states obtained from the superposition of two different states do have physical implications \cite{9}.

In \cite{10} Oszmaniec \emph{et al.} dealt with an interesting superposition problem. Let $\alpha, \beta$ be two given nonzero complex numbers
satisfying $|\alpha|^2 + |\beta|^2 = 1$, and $\mathcal{H}$ a Hilbert space with
dim $\mathcal{H} \geq 2$. Denote $\rho_\psi=|\psi\rangle\langle\psi|$
($\rho_\phi=|\phi\rangle\langle\phi|$) for any pure states $|\psi\rangle$ ($|\phi\rangle$) in $\mathcal{H}$. The problem is whether there is a non-zero trace-nonincreasing completely positive operation $\Lambda _{(\alpha,\beta)}: \mathcal{H}^{\otimes2}\rightarrow \mathcal{H}$: $\Lambda _{(\alpha,\beta)} (\rho_\psi \otimes \rho_\phi )\propto |\Psi\rangle\langle\Psi|$, where $|\Psi\rangle= \alpha|\psi\rangle +\beta e^{i\theta} |\phi\rangle$, the phase $\theta$  may depend on the input pure states, $\Lambda (\rho)=\sum_k M_k\rho M^\dag_k$ for any state $\rho$, $\sum M^\dag_kM_k\leq I$ with $I$ the identity. Here, the postselection is allowed. The desired output state may be obtained probabilistically, depending on outcome of some measurements on the disregarded register.


This no superposition result  is similar to many other no-go theories, such as the no-cloning theorem \cite{11,12,13}, and the recent no-masking theorem \cite{14,15,16}, which both characterize an event or operation that is feasible in the classical world but not in the quantum world. However, there are still difference between them, for two unknown random pure states, even the success probability of superposing  is allowed to be zero, there is still no quantum operation to superpose them with non-zero probability \cite{10}. This result gives  another no-go theorem in quantum theory.
Since the probability of successful superposition of some input state pairs is allowed to be zero under post-selection, an interesting question is
to characterize the feasible set of quantum  state pairs which are of zero success probability of superposition.

 In this paper, we consider the  zero success probability superposition quantum operation be still valid.
We further prove there are enough state pairs can not be superposed with non-zero probabilities, for example, the state pairs constitute an uncountable set.
Our result is then partially answer a conjecture in \cite{Md,MX,S},  that is, the state pairs with nonzero success probabilities of superposition are on a restricted set or satisfy some constraints. This paper is organized as follows, in Section \ref{Preliminary}, we first provide some notations and definitions will be used in the following sections, in Section \ref{super}, we characterize the superposition set for random unknown state with respect to a known state undergoing a trace-nonincreasing completely positive operations. In Section\ref{qudit}, we illustrate the superposable set in qudit system, and  Section\ref{conclusion} is the conclusion and prospect.

\section{Preliminary knowledge}\label{Preliminary}
We call $\Delta$ \emph{a complete superposable states set with respect to the $CP$ operation $\Lambda _{(\alpha,\beta)}$} if for arbitrary chosen $|\psi\rangle,\,|\phi\rangle\in \Delta$, there are the \emph{CP} operation  $\Lambda _{(\alpha,\beta)}$ which implement the superposition with non-zero success probability according to
the definition of superpositions of  $\mathbb{P}_\psi$ and
$\mathbb{P}_\phi$  in \cite{10}. If the volume measurement $M(\mathcal{H})$ of the set of all state points in the space $ \mathcal {H} $ is a finite positive number, is the volume measurement $M(\Delta) $ of $ \Delta $ greater than zero?  Obviously $M(\Delta)<M(\mathcal{H})$, otherwise, it indicates that the  $\Lambda _{(\alpha,\beta)}$ operations fails in only one zero metric set (this set may have infinite elements), which means the $\Lambda _{(\alpha,\beta)}$  is workable on almost any two unknown states in $\mathcal{H}$.
In order to determine the size of a complete superposable state set $\Delta$, let's consider a special case:
we already know a state $|\phi_0\rangle$ in $\Delta$, obviously, the \emph{CP} operation $\Lambda _{(\alpha,\beta)}$
is workable (superposition) for arbitrary states (including $|\phi_0\rangle$) and $|\phi_0\rangle$,
now we consider the set $\Omega$ which of all states that can be superposed with $|\phi_0\rangle$ under the action of the $\Lambda _{(\alpha,\beta)}$, then it is obvious that $\Omega\supset \Delta$, namely $M(\Omega)\geq M(\Delta)$. In other words, compared with the study on superposition of two unknown states that can be linearly superposed with non-zero probability, the study on superposition of unknown states with a known state, the set has a wider scope in mathematics. We have obtained the following results $M(\Omega)=0$ in Theorem 1,  therefore $M(\Delta)=0$.

For given $(\alpha, \beta)$, we denote that a set $\Omega\subseteq \mathcal{H}$ is a superposable states set with respect to a known state $|\phi_0\rangle \in \Omega $ and the \emph{CP} map $\Lambda _{(\alpha,\beta)}$ if the $\Lambda _{(\alpha,\beta)} (\rho_{\psi_s }\otimes \rho_0 ) \propto |\Psi\rangle\langle\Psi|$ for all $|\psi_s\rangle \in \Omega$, where $\rho_0=|\phi_0\rangle\langle\phi_0|$, $\rho_{\psi_s}=|\psi_s\rangle\langle\psi_s|$ and $|\Psi\rangle=\alpha|\psi_s\rangle +\beta e^{i\theta(s)}|\phi_0\rangle$. The phase $\theta(s)$  may depend on $|\psi_s\rangle, |\psi_0\rangle$ ( phase $\theta(s)\in [0,2\pi)$ that may in general depend on the
input states). The above definition comes entirely from \cite{10}.

Notice that the impossibility of superposition of two qubit states implies the impossibility of the superposition of two high dimensional states. While all qubit pure states correspond to the points on the spherical circle one by one on the Bloch sphere, the geometric area on the Bloch sphere can be used to measure the size of the set of the superposable states. In the following, we will focus on the superposition of quantum states in qubit systems.

\section{The superposable sets with respect to a  known qubit state.}\label{super}

Without loss of generality, we assume the known state to be $\rho_0=|0\rangle\langle0|$.
An arbitrary pure qubit state $|\psi\rangle$ can be written as
$$
|\psi\rangle=\cos\frac{x}{2}|0\rangle+e^{-yi}\sin\frac{x}{2}|1\rangle,
$$
where $x\in [0,\pi]$ and $y\in[0,2\pi)$. Set $\cos x=Z$, $\sin x\cos y=X$ and $\sin x\sin y=Y$, one has $X^2+Y^2+Z^2=1$. $(X,Y,Z)$ is the point on the Bloch sphere on which we can define an ``area" measure for a set of qubit states. The total area of all the qubit states is ${4\pi}$.


We search for all states $\rho_\psi$ such that
$\Lambda_{(\alpha,\beta)}(\rho_\psi\otimes\rho_0)= \sum_k M_k(\rho_\psi\otimes\rho_0) M^\dag_k$, $\sum_k M^\dag_kM_k\leq I,$  we only need to consider the conditions holds for all $k$ ,
$M_k(\rho_\psi\otimes\rho_0) M^\dag_k\varpropto |\Psi\rangle\langle\Psi|$ [see \cite{10} for detailed reasons].
Consequently, it is enough to consider only $CP$ maps that
have one operator $M_k$
\begin{eqnarray}\label{3}
H=M_k(\rho_\psi\otimes\rho_0) M^\dag_k-\lambda|\Psi\rangle\langle\Psi|=0,~~~ \lambda>0,
\end{eqnarray}
where $|\Psi\rangle=\alpha|\psi\rangle+\beta e^{\theta i}|0\rangle$. $M_k$ can be expressed in general,
\begin{eqnarray}\label{2}
M_k=\left(\begin{array}{cccc}
a_{11} & a_{12} & a_{13} & a_{14} \\
a_{21} & a_{22} & a_{23} & a_{24} \\ \end{array} \right),
\end{eqnarray}
where $a_{ij}$ are some complex constants [see\cite{10} and its Supplemental Material]. Then the elements $H_{ij}$ of the matrix $H$ must satisfy the following equations,
\begin{equation}\label{fkj}
\begin{array}{ll}
H_{11}(x,y)=&|a_{11}\cos(\frac{x}{2})+ a_{13}\sin(\frac{x}{2})e^{-yi}|^2\\[1mm]
&- \lambda |\alpha\cos(\frac{x}{2})+\beta e^{\theta i}|^2=0,\\[2mm]
H_{12}(x,y)=&(a_{11}\cos(\frac{x}{2})+a_{13}\sin(\frac{x}{2})e^{-yi})\\[1mm]
&*(a_{21}\cos(\frac{x}{2})+a_{23}\sin(\frac{x}{2})e^{-yi})^\dag\\[1mm]
&-\lambda(\alpha\cos(\frac{x}{2})+\beta e^{\theta i}) (\alpha\sin(\frac{x}{2})e^{-yi})^\dag=0,\\[2mm]
H_{22}(x,y)=&|a_{21}\cos(\frac{x}{2})+a_{23}\sin(\frac{x}{2})e^{-yi}|^2\\[1mm]
&-\lambda|\alpha\sin(\frac{x}{2})e^{-yi}|^2=0,
\end{array}
\end{equation}
where $\lambda>0$, $\lambda$ and $\theta$ may vary with $(x,y)$.

\emph{Theorem 1.}  The superposable states set $\Omega$  with respect to a known state $|\phi_0\rangle \in \Omega $ in terms of trace-nonincreasing completely positive operations are located on finite circles on the Bloch sphere.

\emph{Proof.} Since \emph{CP} mapping $\Lambda _{(\alpha,\beta)} $ can superpose $|0\rangle$ and $|0\rangle$ with non-zero probability obviously, taking $x=0$ in $H_{22}(x,y)$ we obtain
\begin{equation} \label{4}
a_{21}=0, ~\mathrm{and~then},~ \forall x\neq0,~ \lambda=\frac{|a_{23}|^2}{|\alpha|^2}.
\end{equation}

Generally, as $\alpha,\beta\neq0$ and $\lambda>0$, one has $a_{23}\neq0$. Substitute (\ref{4}) into $H_{12}(x,y)$, we obtain $\forall x\neq0$
\begin{equation}\label{5}
(a_{11}-a_{23})\cos(\frac{x}{2})+a_{13}e^{-yi}\sin(\frac{x}{2})
=\frac{a_{23}\beta}{\alpha}e^{\theta i}.
\end{equation}
Denote $a_{11}-a_{23}=r_1e^{\gamma_1 i}$, $a_{13}=r_2e^{\gamma_2i}$ and $\gamma=\gamma_2-\gamma_1$.
Due to the arbitrariness of $\theta$, (\ref{5}) holds if and only if
\begin{equation}\label{6}
|r_1\cos(\frac{x}{2})+r_2e^{(-y+\gamma)i}\sin(\frac{x}{2})|=|\frac{a_{23}\beta}{\alpha}|,
\end{equation}
namely
\begin{eqnarray}\label{7}
A\cos x+B\sin x\cos y+C\sin x\sin y+D=0,
\end{eqnarray}
where
\begin{eqnarray}\label{8}
A=\frac{r_1^2-r_2^2}{2},~ B=r_1r_2\cos(\gamma),~~~~~~\nonumber \\
C=r_1r_2\sin(\gamma),~
D=\frac{r_1^2+r_2^2}{2}-|\frac{a_{23}\beta}{\alpha}|^2.\end{eqnarray}

Obviously, $A$, $B$ and $C$ are not all zero, otherwise, $r_1=r_2=0$ and (\ref{6}) can not be satisfied. Since (\ref{5}) also leads to $H_{11}(x,y) = 0$, so $H=0$. Hence, the equation (\ref{7}) just represents a circle formed by the intersection of a plane and a Bloch sphere.

Note that $M_k$ is essentially finite, so spherical circles are finite, the theorem is proved. $\blacksquare$

Theorem 1 is actually true for all the \emph{CP} maps in the sense of superposition defined in \cite{10}.
We see that the complete superposable set $\Delta$ is a subset of the superposable set $\Omega$ with respect to a known state $\rho_0\in\Delta\subseteq\Omega$ and the same $\Lambda _{(\alpha,\beta)}$. Since the area measure of finite circles on Bloch sphere is zero, $M(\Delta)=M(\Omega)=0$.  One of the important result of Theorem 1 is to determine the size of  a completely superposable set $\Delta$, namely

\emph{Corollary.} The measure of any completely superposable set $\Delta$  is zero.

By corollary, we have the results, in consistent with the results given in \cite{10}, no trace-nonincreasing completely positive operations can superpose two unknown pure quantum states with non-zero probability.

\emph{Theorem 2.} Any states located at a circle on Bloch sphere are superposable with a known  state under trace-nonincreasing completely positive operations.

\emph{Proof.} Denote the known state $|\phi_0\rangle=|0\rangle$ at \emph{Z}-axis direction. The equation satisfied by any states at a circle on Bloch sphere has the form,
\begin{eqnarray}\label{9}
Z\cos\mu+ X\sin\mu \cos \nu +Y\sin \mu\sin \nu +c=0,
\end{eqnarray}
where $\mu,\nu$ are constants, $Z=\cos x$, $X=\sin x\cos y$ and $Y=\sin x\sin y$. The point $(X,Y,Z)$ corresponds to the state $|\psi\rangle_s=\cos\frac{x}{2}|0\rangle+e^{-yi}\sin\frac{x}{2}|1\rangle$.

We only need to construct a \emph{CP} map $\widetilde{\Lambda} _{(\alpha,\beta)} $ such that the states $|\psi\rangle_s$ in (9) and $|0\rangle$ are superposed.
Choosing ${\alpha}=\sqrt{\frac{1+\cos\mu}{2+\cos\mu-c}}$ and $\beta=\sqrt{\frac{1-c}{2+\cos\mu-c}}$, let
\begin{eqnarray}
M_0=\left(\begin{array}{cccc}
0 & 0 & \sin(\frac{\mu}{2})e^{\nu i} & 0 \\
0 & 0 & -\cos(\frac{\mu}{2})& 0 \\ \end{array} \right ),
\end{eqnarray}
$\theta=-\arctan(\cot\frac{\mu}{2}\cot\frac{x}{2}\csc(\nu-y)+\cot(\nu-y))-\frac{\pi}{2}$ and $\lambda=\frac{1+\cos\mu}{2|\alpha|^2}$ in (\ref{3}). By direct calculation we obtain \begin{eqnarray}
H=0,~~~|\Psi\rangle=\sin\frac{\mu}{2}|0\rangle+e^{(\nu+\pi)i}\cos\frac{\mu}{2}|1\rangle.
\end{eqnarray}
As $\mathrm{Tr}(M^\dag_0M_0(\rho_s\otimes|0\rangle\langle0|))=\sin^2(\frac{x}{2})$, the probability of success is not zero.

 Since  $ M^\dag_0M_0\leqslant I$, thus the linear operator  $\widetilde{\Lambda}_{(\alpha,\beta)}$ we constructed is completely positive without increasing trace. $\blacksquare$

  Theorem 2 is not inverse proposition of Theorem 1.  In fact,  Theorem 2 is only a sufficient condition, which indicates that for any spherical circle and a known state point, no matter whether the known state point is on the spherical circle or not,  we can have the $CP$ operator, so that the points on the spherical circle and the known state points can be superposed under the definition of superposition in reference\cite{10}. From the proof of Theorem 2, we can see that there is a physical operation that superposes any state in the plane with a fixed state to generate a superposition state perpendicular to the plane with a non-zero probability.   Let us take plane $X + Y + Z + c = 0$ as an example, see Fig. 1. In Fig. 1, $|\Psi\rangle$ is the output state point $(-\frac{1}{\sqrt{3}},-\frac{1}{\sqrt{3}},-\frac{1}{\sqrt{3}})$. For any given state on a spherical circle, there is a \emph{CP} operator $\widetilde{\Lambda}_{(\alpha, \beta)}$ which maps it to the state $|\Psi\rangle$. Namely, the information encoded in the states on each spherical circle can be deleted with certain probability.
Because any state must be on a parallel ring on the Bloch sphere, we can conclude that the information of all the states can be deleted probabilistically by the \emph{CP} operator $\widetilde{\Lambda}_{(\alpha, \beta)}$.

For any state $|\psi\rangle=\cos\frac{x}{2}|0\rangle+e^{-yi}\sin\frac{x}{2}|1\rangle$ and the corresponding operator $\widetilde{\Lambda}_{(\alpha, \beta)}$, the probability of deleting the information (transforming to the fixed state) is $\sin^2(\frac{x}{2})$ under superposition. Consider the set of states on a ring characterized by $Z+\frac{1}{2}=0$, namely, $|\psi\rangle_s=\frac{{1}}{2}|0\rangle+e^{-yi}\frac{\sqrt{3}}{2}|1\rangle$. From Theorem 2 we have the corresponding operator $\widetilde{\Lambda}_{(\alpha,\beta)}$ with $\alpha=\sqrt{\frac{4}{5}}$, $\beta=\sqrt{\frac{1}{5}}$, $\lambda=\frac{5}{4}$ and $\theta=-\pi$, which superposes any states $|\psi_s\rangle$ and $|0\rangle$ to the state $|1\rangle$ with probability $75\%$ by $M_0$. Although the resulting states do not contain the phase information about the original states, these probabilities reflects the distribution of information. Not surprisingly, the results are the same as the ones from projective measurements on the state $|\psi_s\rangle$. Here, the so-called post selection of \emph{CP} operation is in fact the result of POVM collapse of high-dimensional states. That is, the probability superposition and the collapse are the two sides of the same object.
\begin{figure}[h]
\scalebox{2.5}{\includegraphics[width=4.0cm]{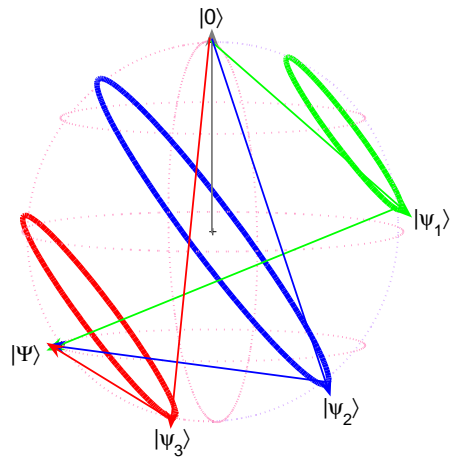}}
\caption{ The green circle for $c=-1.5$, the blue circle for $c=0$, and the red circle for $c=1.3$. $|0\rangle$ is superposed with any states given by the points on the same spherical circle, giving rise to the fixed state $|\Psi\rangle$ with probability $\sin^2(\frac{x}{2})$. $|\psi_1\rangle$, $|\psi_2\rangle$ and $|\psi_3\rangle$ are the states given by the three points on the three different spherical circles, respectively, which give rise to the maximum probability.}
\label{transition}
\end{figure}

The $\widetilde{\Lambda}_{(\alpha, \beta)}$ operator above is not the only one that carries out the superposition for the same spherical circle states. In fact, a  set of ring states can be superposed with a fixed state by infinitely many \emph{CP} operators, maybe with different probabilities and superposition states. For example, we still consider the spherical circle  $Z+\frac{1}{2}=0$,  we choose
\begin{eqnarray}
M=\left(\begin{array}{cccc}
\frac{\sqrt{2}}{2} & 0 & 0 & 0 \\
0 & 0 & -\frac{\sqrt{2}}{2} & 0 \\ \end{array} \right),
\end{eqnarray}
 let $\alpha=-\frac{\sqrt{2}}{2}$, $\beta=\frac{\sqrt{2}}{2}$, and $\theta=0$, the \emph{CP} operator $M$ superposes any states $|\psi_s\rangle$ and $|0\rangle$, still obtain the state $\frac{{1}}{2}|0\rangle+e^{-yi}\frac{\sqrt{3}}{2}|1\rangle$ with probability $50\%$.
 Operator $\widetilde{\Lambda}_{(\alpha, \beta)}$ which convert input states with different relative phase into the states with the same relative phase may be applied to quantum fault-tolerant calculations with phase noises \cite{18,17},  and the \emph{CP} operator $M$ can play the role of ``record'' for the same spherical circular states with probability $50\%$.  Therefore, the superposable set of states with respect to some \emph{CP} operators  may be used to prepare available quantum resources.  A question raised here is that, in order to get available quantum resources, for  $\Omega_1,\Omega_2 \subseteq \mathcal{H}$, and a \emph{CP} operators  superposes any two states from $\Omega_1$ and $\Omega_2$ respectively to the required source state with non-zero probability,  what the characteristics of $\Omega_1$, $\Omega_2$ and the \emph{CP} operators will be.

Is there any intrinsic relationship between the no superposition theorem and other no go results in quantum mechanics?
 We will discuss the compatibility relationship between the no cloning theorem and the no superposition theorem in the next section.  Using the idea of unknown state and a fixed state, for the high-dimensional case we illustrate that any superposition transformation protocols will violate the no-cloning principle for almost all the states.

\section{Some discussions on qudit states.}\label{qudit}
 Similar to using axiomatic model to judge the correctness of the derived theorem model in mathematics, to understand the no-go theorems in quantum theory one should consider the consequences of violating the physical principles \cite{20}. Concerning the no cloning theorem \cite{22,23}, the problem considered is if all pure states can be superposed, resulting in the violation of the principle of probabilistic cloning. Following this idea we further consider to what extent the no cloning principle would be violated as far as the superposition protocol could.

According to our discussions above on the feasibility of probabilistic quantum operation, the success probability for some states can be zero under the condition of allowing for post selections. This means that the set of states allowed to violate the no cloning theorem can be a zero metric set of infinite elements. In other words, even if we cite the infinite violation of the no cloning principle, we cannot say that the \emph{CP} operator ${\Lambda}_{(\alpha, \beta)}$ is no-go. Therefore, the impossibility of quantum operations actually requires to prove that the state set measure which violates the principle is non-zero.

Let $\mathcal{H}_2$ be a two-dimensional subspace of any $n$-dimensional Hilbert space $\mathcal{H}_n$ $(n\geq3)$. Denote $|\psi\rangle=\cos\frac{x}{2}|0\rangle+e^{-yi}\sin\frac{x}{2}|1\rangle$ an arbitrary state in $\mathcal{H}_2$ under orthogonal bases $|0\rangle,|1\rangle$. Let $|2\rangle$ be a pure state in $\mathcal{H}_n$ which is orthogonal to $\mathcal{H}_2$.
Suppose there is a \emph{CP} operator ${\Lambda}_{(\alpha, \beta)}$ such that
\begin{eqnarray}
\Lambda_{(\alpha, \beta)}(|0\rangle,|2\rangle)\rightarrow
|\Psi_0\rangle= \alpha|0\rangle+\beta e^{i\theta_0}|2\rangle,\nonumber\\
\Lambda_{(\alpha, \beta)}(|1\rangle,|2\rangle)\rightarrow
|\Psi_1\rangle= \alpha|1\rangle+\beta e^{i\theta_1}|2\rangle,\nonumber\\
\Lambda_{(\alpha, \beta)}(|\psi\rangle,|2\rangle)\rightarrow|\Psi\rangle~=~~~~~~~~~~~~~~~~~~~~\nonumber\\
\alpha(\cos\frac{x}{2}|0\rangle+e^{-yi}\sin\frac{x}{2}|1\rangle)+\beta e^{i\theta_{\psi}}|2\rangle\end{eqnarray}
with some nonzero probability, where $\theta_0$ and $\theta_1$ are constants, $\theta_{\psi}$ is related to the input state $|\psi\rangle$, $|\alpha|^2+|\beta|^2=1$ and $\alpha,\beta \neq 0$. Since $|0\rangle$, $|1\rangle$ and $|\psi\rangle$ are linearly dependent, $|\Psi_0\rangle$,  $|\Psi_1\rangle$ and $|\Psi\rangle$ must be linearly dependent. Otherwise, the probability cloning principle will be violated, namely, the states can be probabilistically cloned if and only if they are linearly independent \cite{22}.

We only need to calculate the metric of the states $|\psi\rangle$ meeting the requirements. Namely, the rank of the following matrix should be two,
\begin{eqnarray}
[|\Psi_0\rangle,|\Psi_1\rangle,|\Psi\rangle]=\left(\begin{array}{ccc}
\alpha & 0 & \alpha\cos\frac{x}{2} \\
0 & \alpha & \alpha e^{-yi}\sin\frac{x}{2} \\
\beta e^{i\theta_0} & \beta e^{i\theta_1} & \beta e^{i\theta_{\psi}} \\
\end{array} \right ),
\end{eqnarray}
which gives rise to
\begin{eqnarray}
\cos\frac{x}{2}+\sin\frac{x}{2}e^{i(\gamma-y)}=e^{i(\theta_{\psi}-\theta_0)}\Leftrightarrow \nonumber \\
|\cos\frac{x}{2}+\sin\frac{x}{2}e^{i(\gamma-y)}|=1\Leftrightarrow~~\nonumber \\
\cos\gamma\sin x\cos y+\sin\gamma\sin x\sin y=0,
\end{eqnarray}
where $\gamma=\theta_1-\theta_0$.
Equation (15) actually represents a circle on the Bloch sphere with zero area measure.

We analyze the above results: $|0\rangle, |1\rangle, |\psi\rangle$  are linearly dependent,
and $|\Psi_0\rangle,|\Psi_1\rangle,|\Psi\rangle$ are linearly independent unless $|\psi\rangle$  satisfy (15).
In other words, $|\psi\rangle$ takes almost all points on the Bloch sphere, the existence of the protocol $\Lambda_{(\alpha, \beta)}$  can transform linearly
dependent pure states into linearly independent pure states. Obviously,
such protocols can be used to perform the tasks of unambiguous discrimination and probabilistic  cloning of linearly dependent pure states. It is forbidden in no cloning theorem, since the states can be probabilistically cloned if and only if they are linearly independent\cite{22,25,24}.

 Since the area of all the states $|\psi\rangle$ in a two-dimensional space is ${4\pi}$, almost all the states would violate the no-cloning principle in the sense of probability. As the two-dimensional subspace we considered is an arbitrary one, the same results apply to the space $\mathcal{H}_n$. That is, any superposition transformation
protocol given by the \emph{CP} operator works at most on a set of zero area measures, relative to all states (for the area measure of all states in $\mathcal{H}_n$ see \cite{16}). Therefore, the success probability of superposing almost all states is zero. Otherwise, the no cloning principle would be violated.

\section{Conclusion.} \label{conclusion}
In summary,  we have shown that the superposable state set for a known state in terms of trace-nonincreasing completely positive operation is located on some circles on the Bloch sphere, with the state area measure zero. Therefore, there is no workable probabilistic quantum operations which carry out the random superposition. Furthermore, we have shown that any states in a circle on the Bloch sphere are superposable with respect to a fixed state. As a by-product, our constructed \emph{CP} operator may be also applied to probabilistic deletion of phase information. Finally, for the high-dimensional case we have illustrated that any superposition transformation protocol operations will violate the no-cloning principle for almost all the states. Our results may promote a deeper understanding and useful application of quantum state superposition and collapse.

\bigskip
\emph {Acknowledgments}
This work is supported by the National Natural Science Foundation of China (NSFC) under Grant No.(12065021,12175147,12075159,12171044); Beijing Natural Science Foundation (Grant No. Z190005); the Academician Innovation Platform of Hainan Province; Shenzhen Institute for Quantum Science and Engineering, Southern University of Science and Technology (No. SIQSE202001).


\begin{thebibliography}{28}


\bibitem{1} B. Anderson and M. A. Kasevich, Science 282, 1686(1998).
\bibitem{2} D. J. Wineland, Rev. Mod. Phys. 85, 1103 (2013).

\bibitem{3} R. Horodecki, P. Horodecki, M. Horodecki and K. Horodecki,
Rev. Mod. Phys. 81, 865 (2009).

\bibitem{4} T. Baumgratz, M. Cramer, and M. B. Plenio, Phys. Rev. Lett. 113, 140401 (2014).

\bibitem{5} A. Streltsov, U. Singh, H. S. Dhar, M. N. Bera, and G.
Adesso, Phys. Rev. Lett. 115, 020403 (2015).

\bibitem{6} B. Li , C.L. Zhu, X.B. Liang, B.L. Ye and S.M. Fei, Phys. Rev. A 104, 012428 (2021).

\bibitem{Halder} S. Halder and U. Sen, Phys. Rev. A 107, 022413 (2023).

\bibitem{Aubrun}G. Aubrun, L. Lami, C. Palazuelos, and M. Plavala, Phys. Rev. Lett. 128, 160402 (2022).

\bibitem{MLHu}ML. Hu, H. Fan. Sci. China Phys. Mech. Astron. 63, 230322 (2020).

\bibitem{c} C.H. Bennett, G. Brassard, C. Cr\'epeau, R. Jozsa, A. Peres and W.K. Wootters, Phys. Rev. Lett. 70, 1895 (1993).

\bibitem{7} P. W. Shor, SIAM J. Comput. 26, 1484 (1997).

\bibitem{8} S. Aaronson and A. Arkhipov, Theory of Computing 9, 143-252(2013).

\bibitem{H} H. Wang, et al., Phys. Rev. Lett. 123, 250503 (2019).

\bibitem{Y} Y.L Wu, et al., Phys. Rev. Lett. 127, 180501 (2021).

\bibitem{9} U. Alvarez-Rodriguez, M. Sanz, L. Lamata, and E. Solano, Sci. Rep. 5, 11983 (2015).

\bibitem{10} M. Oszmaniec, A. Grudka, M. Horodecki, A. W\'{o}jcik, Phys. Rev. Lett. 116, 110403 (2016).


\bibitem{11} W. K. Wootters, and W. H. Zurek, Nature 299, 802(1982).

\bibitem{12} N. Gisin and S. Massar, Phys. Rev. Lett. 79, 2153 (1997).

\bibitem{13} A. Lamas-Linares, C. Simon, J. C. Howell, and D. Bouwmeester, Science 296, 712 (2002).

\bibitem{14}K. Modi, A. K. Pati, A. Sen(De), and U. Sen, Phys. Rev. Lett. 120, 230501 (2018).

\bibitem{15}B. Li, S.H. Jiang, X.B. Liang, X. Li-Jost, H. Fan and S.M. Fei
Phys. Rev. A 99, 052343 (2019).

\bibitem{16}X.B. Liang, B. Li, S.M. Fei and H. Fan, Phys. Rev. A 101, 042321 (2020).

\bibitem{MX} M.X. Luo, H.R. Li, H. Lai and X. Wang, Quantum Inf. Process. 16, 297 (2017).

\bibitem{Md} M. Doosti, F. Kianvash, and V. Karimipour, Phys. Rev. A 96, 052318 (2017).

\bibitem{S} S. Dogra, G. Thomas, S. Ghosh, and D. Suter, Phys. Rev. A 97, 052330 (2018).
\bibitem{18}X.B. Liang, B. Li and S.M. Fei, Phys. Rev. A 100, 030304(R) (2019).

\bibitem{17}Z.H. Liu, et al., Phys. Rev. Lett. 126, 170505 (2021).


\bibitem{20} S. Bandyopadhyay, Phys. Rev. A  102, 050202(R) (2020).

\bibitem{23} A. Daffertshofer, A.R. Plastino and A. Plastino, Phys. Rev. Lett. 88, 210601 (2002).

\bibitem{22} L.M. Duan and G.C. Guo, Phys. Rev. Lett. 80, 4999 (1998).

\bibitem{25} A. Chefles, Phys. Lett. A 239, 339 (1998)
 \bibitem{24}L. Hardy and D. D. Song,  Phys. Lett. A 259, 331 (1999).



\end{thebibliography}
\end{document}